\documentclass[]{spie}  %>>> use for US letter paper
\usepackage[]{graphicx}

\title{On-sky single-mode fiber coupling measurements\\ at the Large Binocular Telescope} 

\author{ Andrew Bechter*\supit{a}, Jonathan Crass\supit{a}, Ryan Ketterer\supit{a}, Justin R. Crepp\supit{a}, Robert O. Reynolds\supit{b}, Eric Bechter\supit{a}, Philip Hinz\supit{c}, Fernando Pedichini\supit{d}, Michael Foley\supit{a}, Elliott Runburg\supit{a}, Eleanya E. Onuma\supit{e}, Scott Gaudi\supit{f}, Giuseppina Micela\supit{g}, Isabella Pagano\supit{h}, Charles E. Woodward\supit{i}
\skiplinehalf
\supit{a}Department of Physics, University of Notre Dame, 225 Nieuwland Science Hall, Notre Dame, IN 46556, USA;
\supit{b}Large Binocular Telescope Observatory, University Of Arizona, 933 N. Cherry Ave., Tucson, AZ 85721, USA;
\supit{c}Department of Astronomy/Steward Observatory, University of Arizona, 933 N. Cherry Ave., Tucson, AZ 85721, USA;
\supit{d}INAF Roma, Via di Frascati 33, 00040 Monte Porzio Catone, Italy;
\supit{e}National Aeronautics and Space Administration, 8800 greenbelt road, Greenbelt, MD 20771, USA;
\supit{f}Department of Astronomy, The Ohio State University, 140 W. 18th Avenue, Columbus, OH 43210, USA; 
\supit{g}INAF-Osservatorio Astronomico G. S. Vaiana, Piazza del Parlamento 1, 90134 Palermo, Italy;
\supit{h}INAF-Osservatorio Astrofisico di Catani, via Santa Sofia 78, I-95125 Catania, Italy;
\supit{i}Minnesota Institute for Astrophysics, University of Minnesota, 116 Church St. SE, Minneapolis, MN 55455, USA;
}

\begin{document} 
  \maketitle 

%%%%%%%%%%%%%%%%%%%%%%%%%%%%%%%%%%%%%%%%%%%%%%%%%%%%%%%%%%%%% 
\begin{abstract}

The demonstration of efficient single-mode fiber (SMF) coupling is a key requirement for the development of a compact, ultra-precise radial velocity (RV) spectrograph. iLocater is a next generation instrument for the Large Binocular Telescope (LBT) that uses adaptive optics (AO) to inject starlight into a SMF. In preparation for commissioning iLocater, a prototype SMF injection system was installed and tested at the LBT in the Y-band (0.970-1.065 $\mu$m). This system was designed to verify the capability of the LBT AO system as well as characterize on-sky SMF coupling efficiencies. SMF coupling was measured on stars with variable airmasses, apparent magnitudes, and seeing conditions for six half-nights using the Large Binocular Telescope Interferometer. We present the overall optical and mechanical performance of the SMF injection system, including details of the installation and alignment procedure. A particular emphasis is placed on analyzing the instrument's performance as a function of telescope elevation to inform the final design of the fiber injection system for iLocater.

% Diffraction limited Doppler spectrometers show promise to improve RV precision through the use of adaptive optics (AO). A next generation infrared spectrograph for the Large Binocular Telescope (LBT) named ``iLocater", will use single mode fibers (SMFs) to illuminate the spectrograph optics from $\lambda$ = 0.97-1.30$\mu$m.

% In spring of 2016, a prototype SMF injection system was installed at the LBT to demonstrate, record, and characterize on-sky Y-band (0.970-1.065$\mu$m) SMF coupling efficiencies. Fiber coupling was recorded on target stars with variable air-mass, brightness, spectral type and seeing conditions for six half-nights from April 13th to 18th 2016. We present a preliminary analysis of the fiber coupling results with simultaneously recorded optical and mechanical diagnostic data. This paper details how the instrument was installed, the procedure for aligning the instrument to the optical axis of the telescope, and the overall optical and mechanical performance of the SMF injection system. A specific focus was placed on analyzing the instrument's performance as a function of telescope elevation.       
\end{abstract}


%%%%%%%%%%%%%%%%%%%%%%%%%%%%%%%%%%%%%%%%%%%%%%%%%%%%%%%%%%%%%
\section{INTRODUCTION}
\label{intro}

iLocater is a high resolution, fiber-fed radial velocity (RV) spectrograph under development for the Large Binocular Telescope (LBT). Single-mode fibers (SMFs) will be used for the instrument as they provide a unique illumination source which mitigates modal noise and are capable of increasing the precision of RV measurements beyond that typically achieved by using multi-mode fibers (MMFs) systems\cite{Jovanovic}. The instrument will use a fiber injection system to couple light from each 8.4 meter diameter telescope primary mirror into a SMF. The requirements for efficient SMF coupling are a stable optical beam with low residual wavefront errors that matches the mode field diameter (MFD)\footnote{The MFD is defined as the 1/$e^2$ diameter of a Gaussian profile} of the fiber. Due to atmospheric turbulence, large ground-based telescopes require the use of adaptive optics (AO) to provide the necessary wavefront correction for SMF coupling. 

To date, efficient SMF coupling using AO had never been demonstrated at the LBT. As part of iLocater's development, it was necessary to demonstrate the principle of SMF coupling using the telescope AO system prior to full instrument commissioning. In order to verify this capability, an inexpensive prototype fiber injection system was installed at the LBT to measure and characterize SMF coupling using the telescope AO system. The Large Binocular Telescope Interferometer (LBTI) wavefront sensor (WFS) and the DX adaptive secondary of the LBT were used to provide a corrected input to the system\cite{Hinz}.

The prototype system was designed to demonstrate SMF coupling efficiency (fraction of starlight) in the Y-band (0.970-1.065$\mu$m) and measure characteristics of the AO corrected beam (stability, FWHM etc). By obtaining SMF coupling measurements in advance of iLocater's commissioning date, the principle technology development and estimated throughput required for iLocater were verified. It is important to note that the prototype system only operated in the Y-band while iLocater will work in the Y- and J-bands simultaneously. Thus, additional fiber coupling tests across full wavelength band ($\lambda$=0.97 - 1.30$\mu$m) are required to develop a complete understanding of the throughput. An informed estimate however, can be made by extrapolating the Y-band measurements into the J-band. For details about the full iLocater instrument, see Crepp at al. 2016\cite{Crepp2016}.

In this paper we discuss the techniques used to couple the telescope beam to the SMF and present the optical and mechanical performance of the SMF injection system. Section \ref{instrument} provides an overview of the 2016 fiber injection system with an updated prescription from the previously published design in Bechter et al. 2015\cite{BechterA}. Section \ref{fiberalign} discusses methods used to align the fiber and the time taken to couple to a target star. Sections \ref{optical} and \ref{mechanical} present measurements of the optical and mechanical performance respectively, with a focus on the instrument's atmospheric dispersion corrector (ADC) performance and the effects of fiber stage flexure. Section \ref{conc} gives a summary of the on-sky results and plans for future work. 

%%%%%%%%%%%%%%%%%%%%%%%%%%%%%%%%%%%%%%%%%%%%%%%%%%%%%%%%%%%%%
\section{Instrument Overview} 
\label{instrument}
The fiber injection system was installed and aligned on the DX (right) side of the LBTI at the central bent Gregorian instrument port. The system was mounted on the V-SHARK structure which has a removable swing arm consisting of a beamsplitter and fold mirror used to divert the F/15 telescope beam toward the fiber coupling optical board. The mounting structure was designed and installed by the Instituto Nazionale di Astrofisica (INAF) for use with V-SHARK, a next generation instrument for the LBT\cite{VSHARK}. Daytime alignment was conducted during allocated engineering time from April 1st-12th and on-sky measurements were taken over six half-nights of observations from April 13th-18th 2016.

\begin{figure}
   \begin{center}
   \begin{tabular}{c}
   \includegraphics[height=7cm]{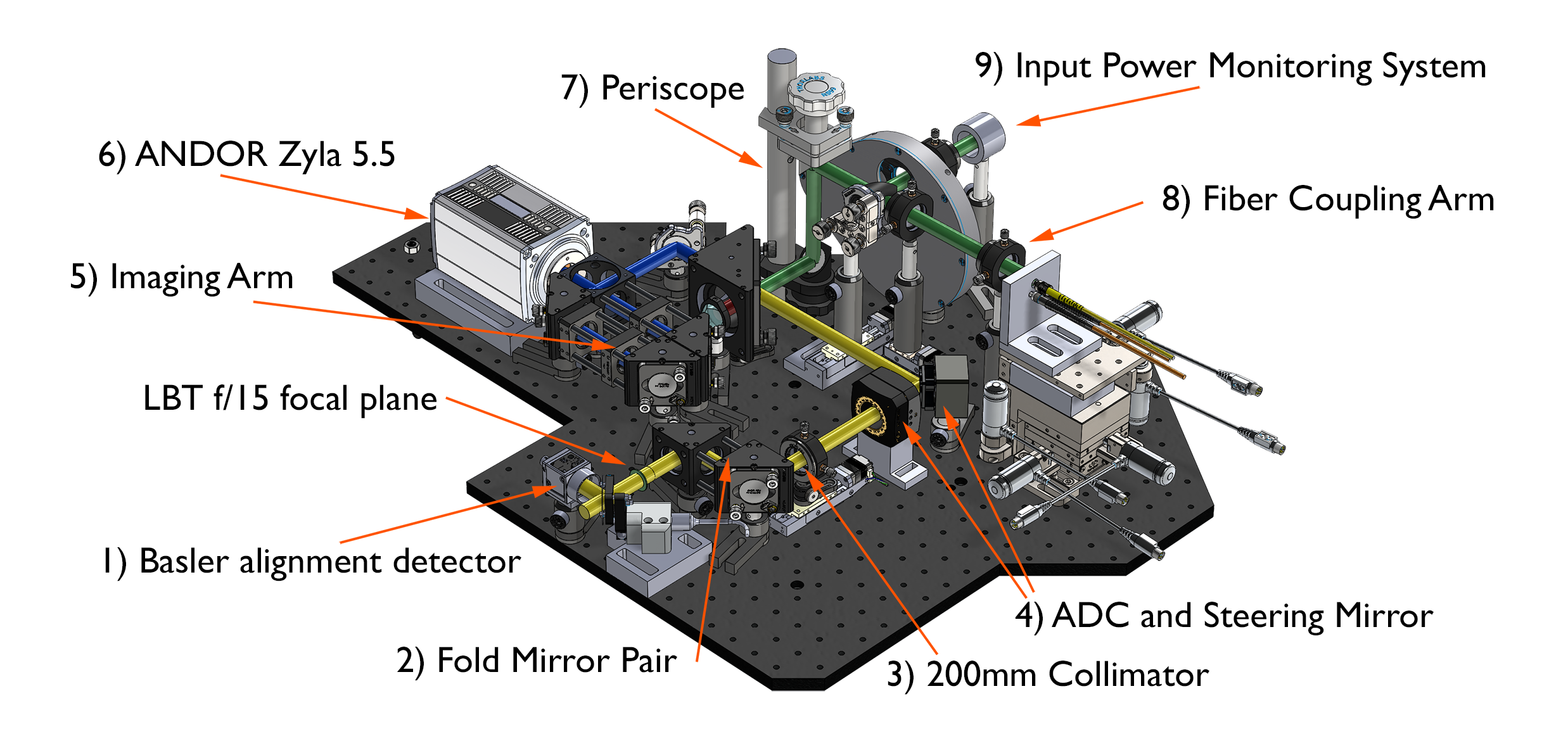}
   \end{tabular}
   \end{center}
   \caption{A 3D CAD rendering showing the full instrument design from Bechter et al. 2015. The full beam path is shown, including all optical components and optomechanics, with common optics in yellow, fiber coupling optics in green, and imaging optics in blue. A Basler alignment camera (1) was used for coarse target acquisition, and an ANDOR (6) detector was used for fine alignment and AO performance monitoring. A third detector was initially installed in place of the fiber channel (7,8,9) for alignment purposes. }
   { \label{overview} }
   \end{figure}

The schematic design of the fiber injection system (Figure \ref{overview}) can be divided into three sections: common optics (yellow), a fiber coupling arm (green), and an imaging arm (blue). The incident AO corrected F/15 telescope beam entered at the bottom of the figure and is first diverted to a target acquisition detector (1) by deploying a fold mirror (2). The WFS on LBTI was moved to adjust the location of the target star to be co-incident with a known reference pixel marking the location of a correctly aligned beam. This detector (Basler acA3800-90um) was selected for its relatively large chip size corresponding to a field of view of about 15'' which aided in acquiring the target star. After retracting the fold mirror, the target was collimated by an achromatic doublet (3), and transmitted through the ADC (4)\footnote{The prism wedge angle and model (Thorlabs PS810-B) were updated since the 2015 design} in the null position before being split into a reflected and transmitted beam by a 2'' diameter laser line beamsplitter\footnote{The beamsplitter model (Newport 20QM20HM.15) and diameter were updated since the 2015 design}. 

The transmitted beam (0.750-0.970 $\mu$m) was propagated through the beamsplitter into an imaging channel (blue), used to assist in fiber alignment, monitoring AO performance, and tracking target star position. Fine adjustments of the beam position were made by the WFS to align the beam with a reference pixel on the ANDOR (Zyla 5.5) CMOS detector (6). This procedure ensured the target was aligned on-axis in the fiber injection optical system. Once the system was aligned using the ADC in the null position, the ADC prisms were rotated to the correct orientation for atmospheric dispersion using a look up table. A steering mirror (4) was used to correct for angular deviations of the optical beam caused by the rotating prisms by adjusting the tip/tilt of the mirror until the optical beam was re-aligned to the reference pixel on the ANDOR detector. This process allowed for ADC correction while maintaining on-axis alignment through the system. For details about the ADC performance on-sky see Section \ref{sec_ADC}.

The reflected beam ($\lambda$ \textgreater 0.950 $\mu$m) entered the fiber coupling arm (green) and a steel periscope (1'' diameter) raised the beam height from $\sim$60mm to 160mm to match the height of the fiber ferrules mounted on the 5-axis fiber stage (8). The beam was demagnified by a lens pair to match the mode field diameter (MFD) of the SMF, while not exceeding the fiber's numerical aperture. The incident Y-band power was measured by deploying a fold mirror into the fiber injection arm to direct the beam through a Mauna Kea Y-band filter (0.97-1.065$\mu$m) before being focused onto a photodiode (9) (Newport 818-IG). The difference in transmission between the input power monitoring optics and fiber injection optics was calibrated using a 1064nm laser and a filtered white light source prior to the observing run. This allowed for any offsets in transmission between the optical paths to the fiber tip and incident power sensor to be calibrated for fiber coupling efficiency calculations. 

The fiber stage (Newport M-562-XYZ) was equipped with three optical fibers: one SMF (Fibercore SM980-5.8/125) and two MMFs from Thorlabs (50$\mu$m core and 200$\mu$m core). The three fibers were installed in a triangular pattern with each fiber in its own ferrule. The MMFs were installed as a fallback plan if the conditions were not suitable for AO correction, or if the SMF could not be aligned using the back-illumination technique (see Section \ref{fiberalign}). For this reason the fiber stage provided 12mm of motion in each dimension for fiber alignment and switching between the fibers however, the MMFs were not required during the observing run. See Bechter et al. 2015 for full design discussion of this fiber injection instrument\cite{BechterA}.

An output monitoring system was designed to mount the output fiber ferrules and record the output power in the Y-band. This was measured by collimating the output from the SMF, transmitting through another Mauna Kea Y-band filter, and focusing the beam onto a second photodiode (Newport 818-IG). Mauna Kea filters (Omega Optical) were chosen from the same batch and similar locations on the original bulk substrate to ensure closely matching transmission curves. The filter transmissions were measured in-house at Notre Dame to confirm similar transmission curves (using a filtered supercontinuum source). The transmission of the output system was calibrated by measuring transmission loss through each component individually. Fiber coupling efficiency was quantified by calculating the ratio of output power to input power in the Y-band. The output system also contained a retractable fold mirror used to inject 1064nm laser light backwards through the SMF into the fiber system for alignment purposes. A retroreflector (shown in red in Figure \ref{overview}) was used to back-propagate light from the fiber focal plane to the imaging arm focal plane. This technique is described in further detail in Section \ref{fiberalign}\cite{Bottom}.

All lenses used in the fiber injection system and fiber output system were Thorlabs or Edmund Optics cemented achromatic doublets with NIR coatings. Commercial off-the-shelf (COTS) lenses were used for their wide selection of focal lengths and low cost. However, these lenses had relatively poor surface quality of $\lambda/4$ compared to the other optics in the system, with $\lambda/20$ mirrors, $\lambda/10$ prisms, and a $\lambda/10$ beamsplitter. While achromatic doublets were the best practical choice, this did not allow the full iLocater wavelength range (Y and J) to be tested simultaneously due to chromatic focal shift. A lens pair was used in the imaging arm of the instrument to reduce the instrument footprint and achieve the desired focal length to magnify the F/15 beam appropriately.

%The final (optional) component of the output system was a Meadowlark Polarimeter\footnote{Due to a loss in observing time, the Polarimeter was not tested during this observing run}, intended to monitor the variations in polarization state at the output of the fiber.

%%-----------------------------------------------------------
\subsection{Installation and alignment} 
\label{alignment}

\begin{figure}
   \begin{center}
   \begin{tabular}{c}
   \includegraphics[height=7cm]{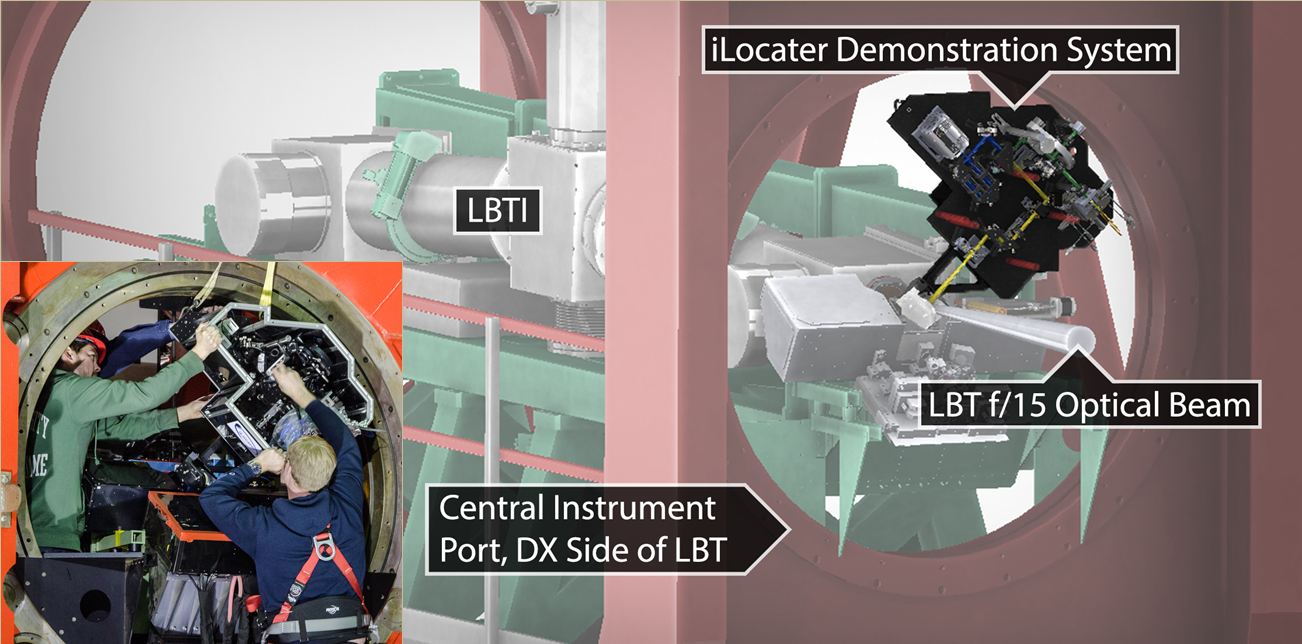}
   \end{tabular}
   \end{center}
   \caption{A 3D CAD rendering showing the location of the SMF injection system as mounted on LBTI. The instrument is located within the central DX instrument port structure shown in red. Incident light from the telescope enters the port and is diverted toward the fiber injection system by a 50/50 beamsplitter and fold mirror in a pick-off arm installed by the INAF V-SHARK team. Inset: Installation of the SMF injection system at the LBT by the iLocater team and LBTO staff.}
   { \label{InstallPhoto} }
   \end{figure}

As part of the installation process, a critical task was aligning the system to the telescope axis at the location of the V-SHARK mounting structure (Figure \ref{InstallPhoto}). Two alignment sources were used for daytime measurements: the first was an on axis alignment source from LBTI (HeNe/Halogen) using a retroreflector installed at the focal plane between the primary and secondary mirror, and the second was the ARGOS on-axis alignment source\cite{ARGOS}. Alignment of the fiber system was accomplished by installing a series of preliminary test-beds which used these alignment sources to measure the optical axis of the telescope and focal plane location prior to mounting the final set of optics. 

The first measurements were taken in March 2016, which located the F/15 focus and the telescope axis on the V-SHARK optical board. The tip/tilt of the V-SHARK pick off optics were adjusted until the optical beam from the telescope was coincident with the location of the optical axis on the instrument board. These measurements provided a coarse position for the fiber injection unit's entrance optics including the fold optics, collimating lens, ADC and steering mirror. A pair of fold optics were mounted in adjustable tip/tilt mounts to refine the alignment of the F/15 axis after installation. As a precaution, a stage was included on the fiber injection system's collimating optic to adjust for focal shift beyond that which the AO system could correct for. 

A second set of measurements was taken by mounting the fiber injection system optical board with a modified set of optical components on April 2nd. The modified set of optics included three detectors and re-imaging optics that were used to identify the telescope axis and focal planes within the fiber injection system. Two of these detectors are shown in Figure \ref{overview}: the Basler detector located at the entrance of the system and the ANDOR detector in the transmission imaging channel. A temporary Basler detector (acA3800-14um) was used in place of the fiber arm. Together these detectors marked three reference points, with one in each optical arm of the instrument, used for reproducing the telescope axis when aligning the final set of optics. The injection system was then removed from the telescope and fully constructed. Final installation and alignment of the unit occurred two days prior to on-sky observation.

%%-----------------------------------------------------------
\section{Fiber Coupling Technique} 
\label{fiberalign}
\begin{figure}
   \begin{center}
   \begin{tabular}{c}
   \includegraphics[height=7cm]{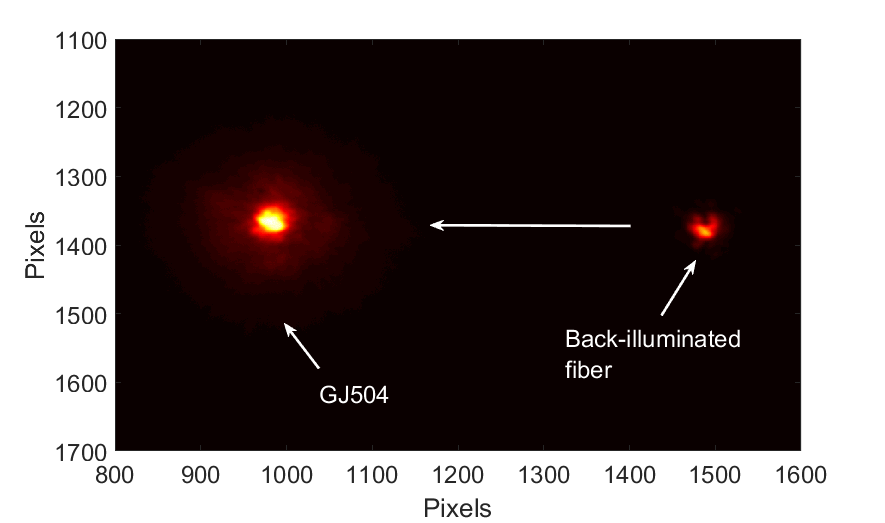}
   \end{tabular}
   \end{center}
   \caption{Image of target start GJ504 and back-illuminated beam on the ANDOR detector. The left PSF is a low order AO-corrected image of the star and the right is the back-illuminated image of the fiber tip using the retroreflector. To achieve fiber coupling, the back-illuminated PSF was positioned on top of of the stellar PSF by driving the fiber stage in the horizontal and vertical directions. After co-aligning the two PSFs, a few percent of coupling efficiency was immediately achieved which allowed for more precise alignment and optimization using the output power sensor.}
   {\label{Back-ill} }
   \end{figure}

The method used for single-mode fiber alignment is a `back-illumination' technique where laser light is back-propagated through the SMF to aid in alignment of the fiber to the telescope axis. For this process, a fold mirror is deployed in the output power monitoring system to direct laser light from a 1064nm, 20mW, laser into the output end of the SMF. Laser light exits through the fiber's input ferrule and transmits `backwards' through the system by using the retroreflector and onto the ANDOR detector shown in Figure \ref{overview}, producing an image which marks the location of the fiber tip. This procedure allows for coarse alignment between the telescope and fiber by co-aligning the telescope point spread function (PSF) and the back-illuminated fiber image on the ANDOR detector. Figure \ref{Back-ill} shows an example of the back-illuminated image prior to alignment with the input PSF from the telescope. Back-illumination required approximately 1 minute to complete and resulted in a measurable quantity of starlight propagating through the fiber after this process. Fine alignment was achieved by retracting the back-illumination fold mirror and driving the five-axis fiber stage until the Y-band power on the output power sensor was maximized. After all five axes on the stage were positioned initially, the stage usually only required small lateral (x-axis) and vertical (y-axis) motion to maximize throughput for each target. After a realistic maximum coupling efficiency was determined for a typical target based on the input sensor power, the alignment process required $\sim$1-5 minutes to optimize throughput. Using the back-illumination technique to directly identify fiber's position relative to the target PSF prevented the need for extensive searching (by raster scanning or other means), and proved to be extremely time-efficient.  

% The majority of SMF alignment time was spent operating the AO loop and inputting instrument parameters (integration times, frame rates etc).

% The fiber stage was equipped with three optical fibers: one SMF (Fibercore SM980-5.8/125), and two multi-mode fibers (MMFs) from Thorlabs (50$\mu$m core and 200$\mu$m core). The three fibers were installed in a triangular pattern with each fiber in its own ferrule. The MMFs were installed as a fall-back plan if the conditions were not suitable for AO correction, or if the SMF could not be aligned using the back-illumination technique. The latter alignment procedure used either MMF to first identify the PSF location and then used a known distance between the MMF and the SMF cores to align the SMF. For all of the on-sky time, the MMFs were not required for alignment. 

%%-----------------------------------------------------------
\section{Optical Performance} 
\label{optical}
% The fiber injection system was designed to re-image the F/15 beam from the LBT for SMF coupling, imaging on the WFC, guiding with the ANDOR camera, and for input power measurement on the input photodiode. This section discusses the measured optical parameters, limitations of the optical design, and plans for improving the optical performance in the final system design. 

% This section discusses the measured optical parameters of the fiber system, limitations of the optical design, and plans for improving the optical performance in the final system design. 

%%-----------------------------------------------------------
The Basler alignment detector was located at the focal plane of the F/15 telescope beam. A pixel scale of 8.8mas/pixel was calculated by measuring the separation of the two components of 41-Leo on this detector. Given the pixel pitch of 5.5$\mu$m, the plate scale at the native F/15 focal plane was measured to be 1.6mas/$\mu$m which is in agreement with the listed value of 1.664mas/$\mu$m.  

The ANDOR detector was used for alignment of the target star and to record images of the target PSF in time. Frame rates ranging from 20-150Hz were used to record images of the target star depending on the apparent magnitude. The plate scale measured on the ANDOR detector was $\sim$2.87 mas/pixel with a typical broadband ($\lambda$ = 0.75-0.97$\mu$m) core width, $\sigma$, ranging from 3-5 pixels. These images have been used to track the centroid and FWHM of the PSF in order to diagnose variations in fiber coupling efficiency during post processing. 

%%-----------------------------------------------------------
\subsection{SMF coupling} 

\begin{figure}
   \begin{center}
   \begin{tabular}{c}
   \includegraphics[height=7cm]{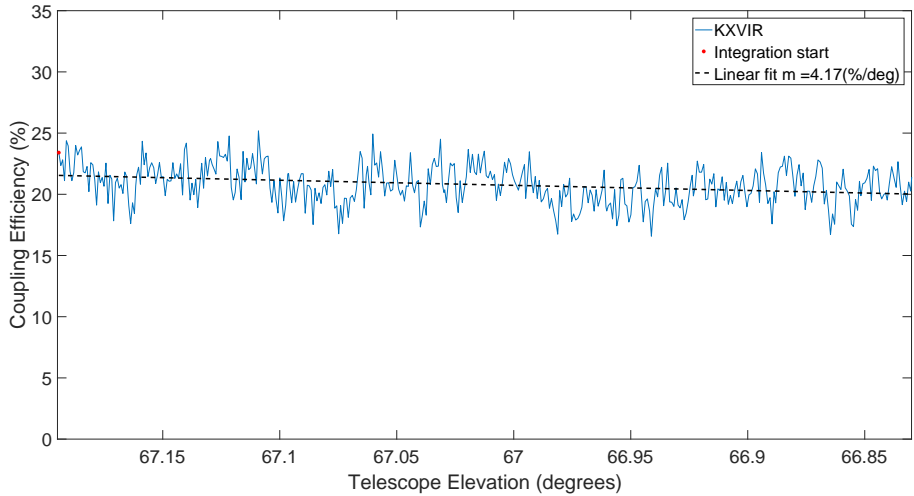}
   \end{tabular}
   \end{center}
   \caption
   {\label{fiber} 
Y-band (0.97-1.065$\mu$m) single-mode fiber coupling efficiency for KX Vir with 400 mode AO system correction and ADC correction.   
}
   \end{figure} 

For each target, the input power was measured for a duration of at least 1 minute by deploying the fold optic in the fiber channel. After recording input power, the fiber was coupled and then the output power was recorded for the desired duration. Typical output measurements lasted 2-5 minutes for ADC calibration tests and initial AO correction tests, with longer durations of 10-30 minutes for extended AO correction and flexure experiments. After measuring the output power, the input power was measured again to check for variations in throughput compared to the first input measurement. Although this method of measuring input/output power has the drawback of asynchronicity, it minimizes the number of optical surfaces before the fiber to retain the highest wavefront quality possible and maximizes the incident flux reaching the fiber.    

Preliminary calculations of fiber coupling efficiency indicate a range of values from $\sim$10-25$\%$ for all targets which ranged from an apparent magnitude of V=1 to V=10. The coupling efficiency from a V-magnitude=7.6 target, KX-Vir (HD 113496), is shown in Figure \ref{fiber}. The fiber coupling data has been down-sampled and plotted against the telescope elevation to highlight variations in performance with changes in telescope elevation. A polynomial fit has been applied to the re-sampled data to look for low frequency losses. The best linear fit shows a loss of about 4$\%$ per degree of elevation change. This loss in fiber coupling is symptomatic of a gradual misalignment of the fiber to the telescope. Causes of the gradual loss in coupling are investigated throughout the remainder of this paper.  
 
During observations, a clear correlation was noted  between AO performance and coupling efficiency with both low- and high-order correction necessary being to achieve efficient coupling. A full analysis, including temporal effects, will be presented in Crass $\&$ Bechter et al. 2016 (in prep.)\cite{CrBe}.

%%-----------------------------------------------------------
\subsection{ADC correction} 
\label{sec_ADC}
\begin{figure}
   \begin{center}
   \begin{tabular}{c}
   \includegraphics[height=7cm]{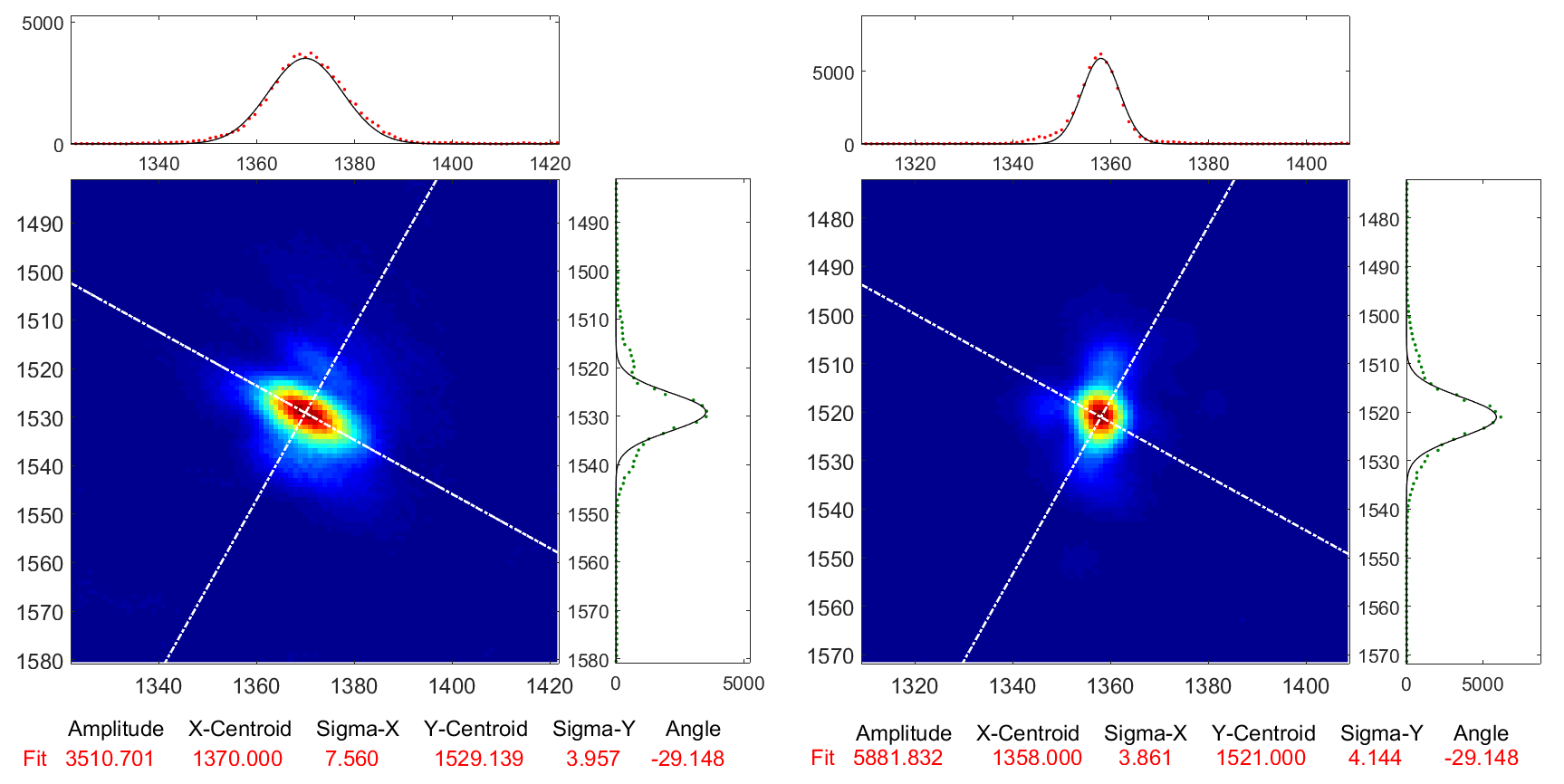}
   \end{tabular}
   \end{center}
   \caption
   {\label{ADC} 
Left: An uncorrected image of 72 Leo at a ZD of 9.8$^{\circ}$. Right: ADC corrected image of 72 Leo at a ZD of 9.7$^{\circ}$. Both are single frames with 0.5 ms integration. 2D Gaussian fit parameters show $\sim$2$\times$ smaller FWHM along the dispersion direction after ADC correction.}
   \end{figure}

An atmospheric dispersion corrector (ADC) was used to optimize single-mode fiber coupling by remapping the dispersed wavelengths from an ellipsoidal PSF to a more desirable circular PSF shown in Figure \ref{ADC}. The ADC used on-sky consists of a pair counter-rotating air spaced singlet wedge prisms in rotation mounts (Newport Conex-AG-PR100P) to correct for varying levels of dispersion. This ADC design created an undesirable output beam angle which varied depending on the rotation angle of the two prisms and required a steering mirror for realignment. These deviations in beam angle prevented active ADC correction when coupling the output beam to a SMF as the varying angle would continuously misalign the telescope beam from the fiber. As a compromise, the ADC was set to a single rotation position throughout a fiber coupling measurement. Due to the lack of active correction, all ADC-specific tests limited fiber coupling efficiency measurements to small elevation changes, corresponding to $\sim$ 2 minutes of elapsed time. It was found that longer fiber coupling integrations are likely to suffer from a lack of ADC correction and faster drop off in coupling efficiency, independent of the AO system stability.  

In order to use the ADC to correct for atmospheric dispersion, the axis of dispersion in the optical system had to be identified. This was measured by fitting a rotated 2D Gaussian to the PSF core and identifying the angle of the long axis. The angle of the dispersion axis was found to be $\sim$30$^{\circ}$ from the plane of the optical board. This measurement was used to set the null positions of the ADC prisms prior to applying a relative rotation. Preliminary measurements indicate the average SMF coupling throughout a 2 minute integration was improved by a factor of $\sim10\%$ at a zenith distance (ZD) \textgreater 15$^{\circ}$ and by 30$\%$ at 9.8$^{\circ}$\textless  ZD \textless 13$^{\circ}$. Accounting for all fiber coupling measurements with the ADC, the efficiency was improved on average by a factor of $\sim25\%$. These measurements indicate a significant improvement in fiber coupling efficiency, and highlight the ADC as a critical component for maximizing SMF coupling efficiency.    

%%-----------------------------------------------------------
\section{Mechanical Performance}
\label{mechanical}
The optical board of the fiber injection system was machined from a Thorlabs aluminum breadboard. A custom designed frame was constructed from aluminum supports and plastic panels, which were mounted around the edge of the board to protect and baffle the instrument. The board was bolted on to the existing V-SHARK backing plate. A lid for the instrument was omitted to prevent turbulent air produced by the ANDOR detector's cooling fan from recirculating into the optical beam path. The optical mounts were constructed from Thorlabs steel posts, aluminum post holders and lens mounts. All lens-positioning stages were Micronix linear stages with optical encoders to keep the lenses positioned changes elevation. As seen in Figure \ref{overview}, the common optics and imaging arm optics are mounted low to the board to prevent flexure. The periscope and one of the focusing lenses in the fiber coupling arm were mounted more than twice as high as the other components in order to match the height of the fiber stage. Due to the range of motion required for switching between fibers, the size of the stage resulted in a significant mass and led to an increased risk of mechanical flexure. The following section discusses the impact of the mechanical design on fiber coupling efficiency.  

\subsection{Mechanical Flexure} 
\label{flex}
\begin{figure}
   \begin{center}
   \begin{tabular}{c}
   \includegraphics[height=7cm]{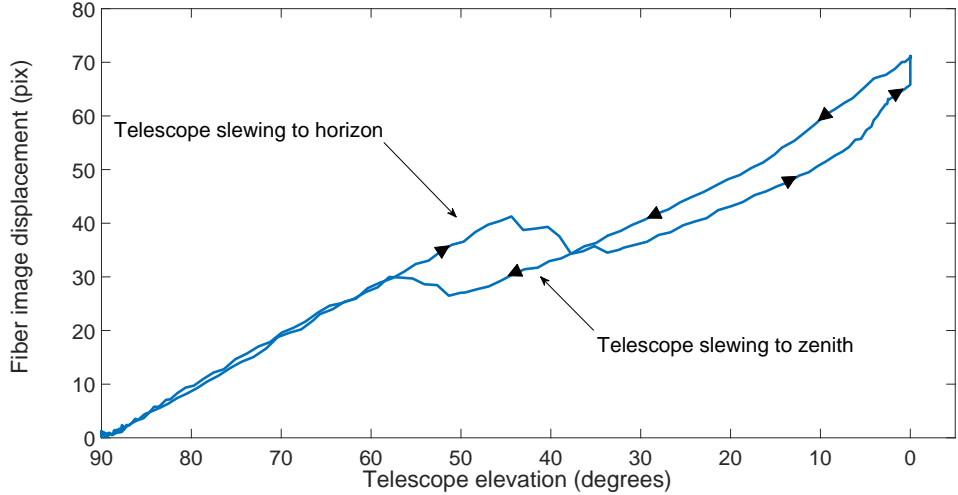}
   \end{tabular}
   \end{center}
   \caption{Displacement of the back-illuminated fiber tip image on the ANDOR detector while the telescope slewed from zenith (90$^{\circ}$ elevation) to horizon (0$^{\circ}$ elevation), and then back to zenith. Flexure measurements indicate as high as 1 pixel displacement/degree with an average of 0.7 pixels/degree across the full telescope range. The loss of symmetry when slewing to zenith and horizon results from the spring-loaded stage design.}
   { \label{flexure} }
   \end{figure}

As shown in Figure \ref{InstallPhoto}, the instrument port holds the optical board perpendicular to the telescope’s axis of rotation (elevation) near to the location of the bent-Gregorian F/15 focus. During an observation, the direction of gravity relative to the instrument varies as the telescope's elevation changes. The resulting flexure on the stage used to position the fibers was measured by recording back-propagating laser light (1064nm) on the ANDOR detector during elevation changes between zenith and horizon. In each frame, the centroid of the back-illuminated fiber image was measured to calculate the displacement of the image centroid relative to the starting position. The fiber stage was identified as the primary cause of flexure by observing significant variations in the stage's optical encoders during telescope elevation change. The displacements from the recorded images were plotted against telescope elevation to determine a correlation of displacement with telescope elevation (Figure \ref{flexure}).

For elevations ranging from 90-60$^{\circ}$, flexure induced displacements showed a linear trend of about 0.7 pixels/degree in both telescope directions (raising and lowering). Elevations between 60-40$^{\circ}$showed an offset `step' depending on the direction of telescope motion. This effect is attributed to the spring loaded mechanism in the 5-axis stage having a different behaviour depending on the direction of the gravity vector. At the lowest elevations, the stage responded similarly to the highest elevations with slight non-linearities being apparent when slewing downwards. Given the motion of the fiber stage due to flexure, the loss in coupling efficiency can be calculated from the overlap integral between the SMF's MFD and the input PSF. Assuming the average effects of mechanical flexure over 5$^{\circ}$ of elevation change, the fiber coupling efficiency would be expected to drop by a factor of $\sim$15-20$\%$ without active fiber alignment due to flexure of the stage.

\subsection{Optical beam drift} 
An on-sky test was devised to simultaneously measure optical drift of the incident PSF and back-illuminated PSF while observing the target star GJ504 for 1 hour. For this test, the PSF of GJ504 and the back-illuminated spot were positioned offset from each other on the ANDOR detector (similar to Figure \ref{Back-ill}). The PSF from the telescope was corrected by the AO system operating with 153 modes. Measuring the telescope PSF's centroid in each frame resulted in an average centroid displacement of 0.06 pixels/degree over the hour of observations. The back-illuminated spot was measured to have a displacement of 0.71 pixels/degree. The measurements indicate GJ504's PSF had an insignificant degree of displacement due to the AO system and re-imaging optics, while the motion of the fiber PSF measurements agree with the flexure values measured in Section \ref{flex} (see Figure \ref{flexure}). This indicates that flexure in the fiber stage resulted in significantly more fiber misalignment than the drift in the centroid of GJ504 governed by the AO system.

\begin{figure}
   \begin{center}
   \begin{tabular}{c}
   \includegraphics[height=7cm]{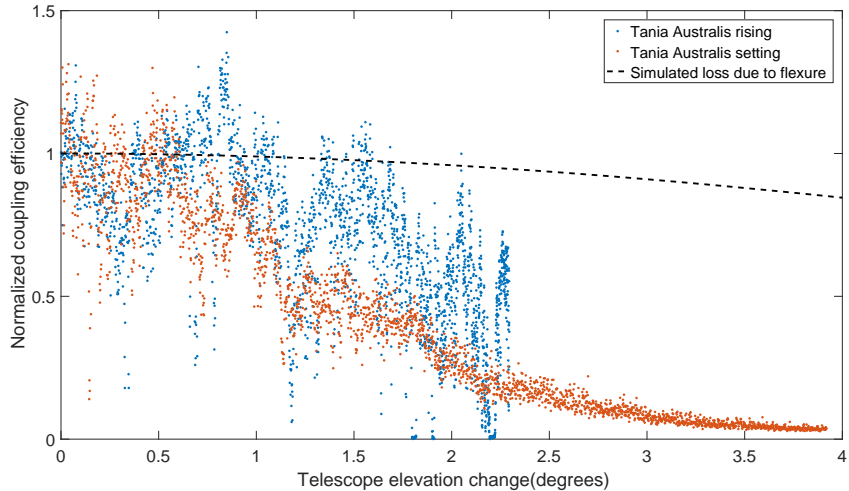}
   \end{tabular}
   \end{center}
   \caption{Normalized coupling efficiencies for Tania Australis recorded on April 18th. The loss in coupling efficiency occurs over a much smaller change in elevation than predicted by flexure effects alone. Simulated loss from flexure predicts $\sim$20 degrees of elevation change before fiber coupling is reduced to 0.}
   { \label{Drift} }
   \end{figure}

Two sets of long-duration fiber coupling data were taken on the target star Tania Australis as shown in Figure \ref{Drift}. This experiment was designed to measure the loss in coupling over large changes in elevation and to assess the realistic maximum exposure times for the final iLocater spectrograph. The data sets were taken on the same night on either side of the star's peak elevation ($\sim$82$^{\circ}$) for 30 minutes each. Both sets of coupling data have been normalized to the first data point to highlight the fractional change in coupling as the target changes elevation. A trace of the simulated coupling loss from flexure effects is over plotted in black based on the calculated flexure in Section \ref{flex} and the GJ504 drift test. It is clear that coupling efficiency in both data sets drops over a smaller elevation change than predicted by flexure effects alone. The likely cause of this drop is a reduction in coupling due to ADC performance as there was no active control after setting the initial dispersion correction values. Further investigation is on-going to explain the slope of the coupling efficiency loss with elevation. 

\section{Summary and Future work}
\label{conc}

The prototype SMF coupling system successfully achieved its primary goal of demonstrating the feasibility of SMF coupling using the AO system at the LBT with fiber coupling efficiencies of $\sim$10-25$\%$. Due to the weather conditions, $\sim$50$\%$ of scheduled on-sky time was lost and thus the targets were limited to a range of V-magnitudes from V=1 to V=10. Fainter targets (V$\sim$16) will be tested during the next phase of SMF testing at the LBT. Improvements in fiber coupling efficiency are expected due to the scheduled LBT AO SOUL upgrade upgrade\cite{SOUL}. Optical beam stability may be improved due to a new instrument location below the fast steering mirrors in LBTI, near the location of the combined focal plane.

The design of the final fiber coupling system to be used in iLocater will include a full set of custom lenses with high surface quality ($\lambda/20$ goal) and minimal chromatic focal shift to be within the tolerances for fiber coupling across iLocater's full wavelength band. Due to the slow F/41.2 beam at the combined focus, the lenses will benefit from a smaller collimated beam diameter with a similar focal length collimating optic to that used after the F/15 focus. A smaller collimated beam will reduce the filling fraction of 25mm diameter optics and allow for easier alignment.

The ADC proved to be critical for improving fiber coupling efficiency, on average improving the fiber coupling efficiency by a factor of $\sim$25$\%$ for all observed targets. Angular deviations from the output of the ADC limited the ZD which could be tested to $\sim$35$^{\circ}$ before the steering mirror retaining ring vignetted the collimated beam. In addition, the COTS singlet prisms limited the wavelength range which could simultaneously be corrected. SMF coupling in the final iLocater instrument will be improved by implementing a zero beam deviation ADC to allow active correction with custom triplet prisms for correction over the full wavelength range ($\lambda$=0.97 - 1.30$\mu$m)\cite{Kopon}. 

Mechanical flexure is a measurable effect with $\sim$0.7 pixels/degree change. This effect is shown to reduce fiber coupling efficiency by $\sim$15-20$\%$ over 5$^{\circ}$ of elevation change. The problem of flexure can be addressed in iLocater's final fiber injection system by improving upon the mechanical design of the fiber stage (avoiding spring-loaded stages), by designing the orientation of the stage to be gravitationally invariant with changes in telescope elevation, or by using high resolution optical encoders to provide feedback to the stage actuators. The fiber stage could also be downsized to include only one fiber ferrule, as the MMFs were not required. In addition, the final system's SMF may include a fiber bundle in which additional fibers are closely packed around a central science fiber. This will allow the surrounding alignment fibers to potentially be back-illuminated simultaneously while taking fiber coupling data. Further investigation is required to determine if active measurement of the back-illuminated images can provide a method of real-time stage adjustment.

Using single-mode fibers for an RV spectrograph offers a new method to facilitate increased resolution and single measurement precision. The successful demonstration of SMF coupling at the LBT marks a major step forward in developing the SMF fiber injection system for use with iLocater. With performance upgrades scheduled for the telescope AO system and the implementation of fiber injection system design improvements, SMF coupling performance at the LBT shows promise to routinely exceed the minimum throughput requirements for iLocater.  

% The prototype system has also allowed for the opportunity to asses the injection system's optical and mechanical design so that upgrades can be implemented before the final system for iLocater is commissioned. The following section provides a summary of the optical and mechanical performance, and provides plans to improve upon the current fiber injection system design. 

\section*{ACKNOWLEDGMENTS}

The iLocater team would like to thank the LBTI team, especially Philip Hinz and Amali Vaz, for support operating the adaptive optics system. We are grateful for the help of LBTO staff and the INAF V-SHARK team, especially Fernando Pedichini, for assistance setting up fiber coupling experiments with LBTI. J. Crepp acknowledges support from the NASA Early Career Fellowship program to develop a fiber-coupling demonstration unit for the LBT. The iLocater team is also grateful for contributions from the Potenziani family and the Wolfe family for their vision and generosity. 

The LBT is an international collaboration among institutions in the United States, Italy and Germany. LBT Corporation partners are: The University of Arizona on behalf of the Arizona university system; Istituto Nazionale di Astrofisica, Italy; LBT Beteiligungsgesellschaft, Germany, representing the Max-Planck Society, the Astrophysical Institute Potsdam, and Heidelberg University; The Ohio State University, and The Research Corporation, on behalf of The University of Notre Dame, University of Minnesota and University of Virginia.

\label{ak}

%%%%%%%%%%%%%%%%%%%%%%%%%%%%%%%%%%%%%%%%%%%%%%%%%%%%%%%%%%%%%
%%%%% References %%%%%

\bibliography{biblistjul13}   %>>>> bibliography data in report.bib
\bibliographystyle{spiebib}   %>>>> makes bibtex use spiebib.bst

\end{document}